\documentclass{article}      
\usepackage{graphicx}
\usepackage{epstopdf}
\title{Elastic nucleon form factors}

\author{M. De Sanctis\\
INFN, Sezione di Roma1, Piazzale Aldo Moro, Roma
 (Italy)\\ 
 and \\
 Universidad Nacional de Colombia, Bogot\`a (Colombia)\\
 M.M. Giannini, E. Santopinto, A. Vassallo,\\
Dipartimento di Fisica dell'Universit\`a di Genova\\
and \\
Istituto Nazionale di Fisica Nucleare, Sezione di Genova, Italy, \\
}

\date{}
\begin{document}

\maketitle  

\begin{abstract}
The relativized hypercentral Constituent Quark Model is used for the calculation 
of the elastic electromagnetic form factors of the nucleon. The results are 
compared with the recent measurements at Jlab.
\end{abstract}

A renewed interest in the electromagnetic form factors of the nucleon has been 
 triggered by the recent results of the Jefferson Laboratory on the
ratio
between the electric and magnetic form factors of the proton \cite{ped}. At
variance
with the expectations, the ratio
deviates
strongly from $1$ and, for $Q^2~\geq~1~(GeV/c)^2$, it
decreases with an almost linear behaviour, pointing towards the
possible
existence of a zero at $Q^2\approx ~8~(GeV/c)^2$.
The main problem is the physical picture emerging from the data, that is the
origin of the decrease of the ratio and of the eventual presence of a
zero in the electric form factor. The first seems to be the manifestation of relativistic 
effects \cite{rel,boffi,rap},
while the possible presence or not of a zero will help to discriminate among the 
different models for the nucleon structure \cite{iac,models,rel,boffi,mds3}. For reviews on this
subject the readers are referred to ~\cite{Gao:2003ag}.
Future experiments at higher $Q^2$ will clarify these points. From the theoretical
point of
view a zero in the proton charge form factor is a challenge for most models of the
internal proton structure. 
In this contribution we report the results of
recent calculations of the elastic nucleon form factors within a semirelativistic version of the 
hypercentral Constituent Quark Model (hCQM) \cite{pl}.

The three quark
hamiltonian for the hCQM is \cite{pl}
\begin{equation}\label{eq:hcqm}
H=\frac{{{\vec{p}}_{\rho}}^{~2}}{2m}÷+
\frac{{{\vec{p}}_{\lambda}}^{~2}}{2m}÷-÷\frac{\tau}{x}÷
+÷\alpha÷x÷+÷H_{hyp}~,
\end{equation}
\noindent where ${\vec{p}}_{\rho}$ and ${\vec{p}}_{\lambda}$ are the conjugate 
momenta of the Jacobi coordinates $\vec{\rho}$ and $\vec{\lambda}$ and $x÷=\sqrt{{\vec{\rho}}^{~2}+
{\vec{\lambda}}^{~2}}$ is the hyperradius. 
The spectrum is described with $\tau~=~4.59$
and $\alpha~=~1.61~fm^{-2}$ and the standard strength of the hyperfine
interaction needed for the $N-\Delta$ mass difference \cite{ik}. 
The $SU(6)_{sf}$ invariant potential is assumed to be  of the 
type linear plus Coulomb-like, a form which is supported not only by the success of the
 the Cornell potential in the meson sector,  but also by recent Lattice QCD calculations \cite{bali} for
$SU(3)_f$ invariant static sources both for the meson and the baryon sectors. 
The model has been used for the prediction of
various physical
quantities of interest, namely the photocouplings \cite{aie},
the electromagnetic transition amplitudes \cite{aie2}, the elastic nucleon
form factors \cite{mds}. The ratio between the electric and magnetic
proton form factors \cite{rap} has been calculated  
 boosting the three quark nucleon states to the Breit frame 
and expanding the matrix elements of the three
quark current up to the first order in the quark momemtum. The non relativistic
calculations predict the value $R=\mu_{p}\frac{G_E^p}{G_M^p}=1$ and introducing the
hyperfine interaction makes no difference ($R=0.99$). However,
the first order relativistic corrections \cite{rap} give rise to a ratio
which significantly deviates from $1$.

We have proposed a semirelativistic version of this model, introducing the relativistic 
kinetic energy:
\begin{equation}
H = \sum_{i=1}^3 \sqrt{{\vec{k}}_{i}^2+{m_i}^2}-\frac{\tau}{x}~
+~\alpha x+H_{hyp},
\end{equation}
\noindent where $\vec{k}$$_i$ 
is the i-th quark 3-momentum in the rest frame, i.e. $\sum_{i=1}^3 ~\vec{k}_i~= ~ 0$ 
and $m_{i}$ are the masses of the constituent quarks.    
This hamiltonian is solved by means of variational tecniques in the nucleon rest 
frame \cite{traini} and the eigenvalues fitted to the reproduction of the existing data 
for the 3 and 4 star resonances.  
The eigenfunctions of Eq. (2) are interpreted as the eigenfunctions of a mass operator in a 
Bakamjian-Thomas construction in a point form dynamics \cite{KP,klink1}.
The nucleon states in the Breit frame are obtained from the rest ones by means of canonical boosts, 
which are purely kinematical and allow the composition of the angular momentum states as in the non 
relativistic case \cite{klink1}.
 
The nucleon electromagnetic form factors can be extracted from the matrix elements of the 
nucleon electromagnetic current according to the formalism described in \cite{klink1,boffi}.
The electromagnetic current is written in impulse approximation, as already done by other 
authors \cite{boffi}
\emph{i.e.} is chosen to be the sum of one-body quark currents \cite{mds,rap}:
\begin{equation}
 J_\mu^{(N)} = \sum_{i=1}^3 e_i~\gamma_\mu(i)
\end{equation}
where $e_i$ is the electric charge of the i-th constituent quark. 
The theoretical predictions are in quite a good agreement with the experimental data.  
Considering that constituent quarks have a finite size \cite{psr}, 
we have introduced constituent quark form factors. The
free parameters in the quark form factors have been fitted to 
the ratio R, the proton magnetic
form factor $G_M^p$, the neutron electric $G_E^n$ and magnetic
$G_M^n$ form factors \cite{mds3}, obtaining a very good reproduction of the data. 
As an example, the results
for the ratio R are shown in Fig. 1. 
\begin{figure}[!ht]
\begin{center}
\includegraphics [width=6in]{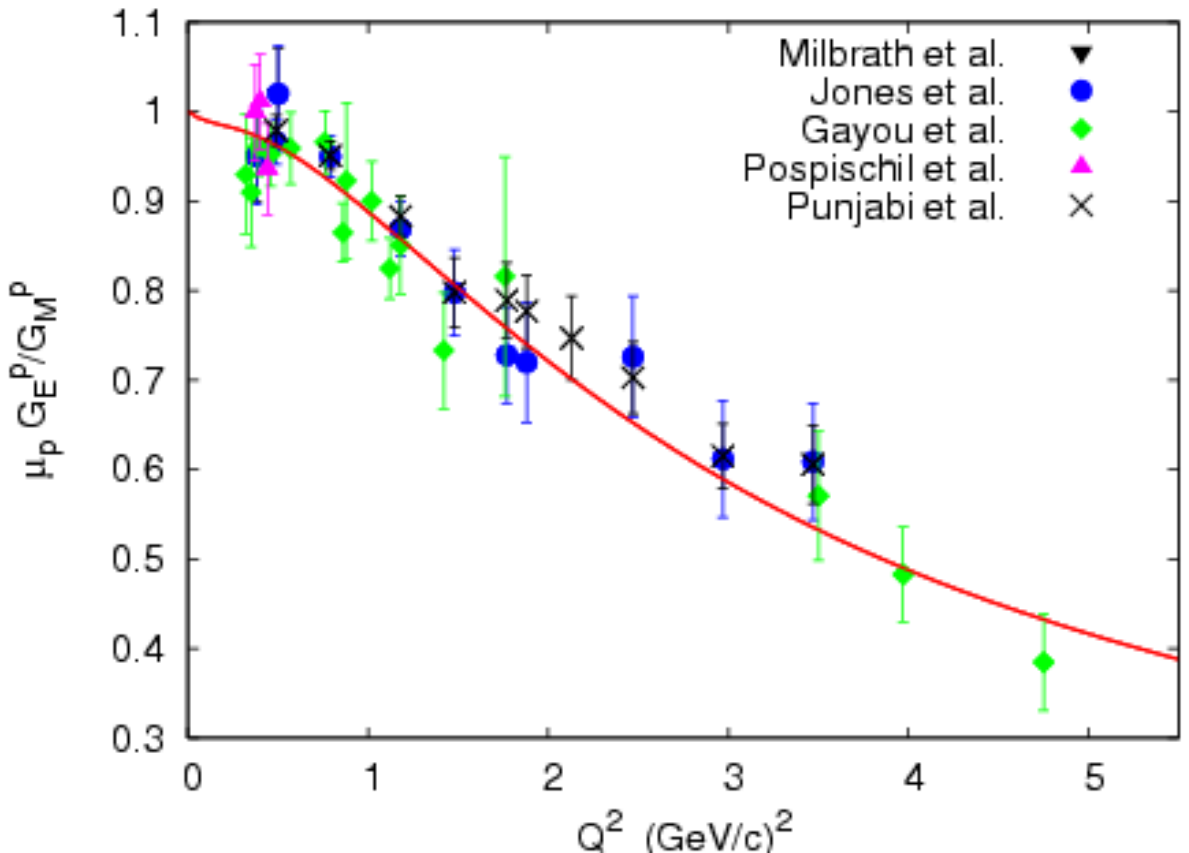}
\caption{
(Color online) 
The ratio $\mu_p G_E^p/G_M^p$ from polarization 
      transfer compared with the semirelativistic hCQM calculation with 
      constituent quark form factors (solid line). 
      The experimental data are taken from~\cite{Milbrath99,ped,Pospischil03,Punjabi:2005}.}
    \label{fig:rap_comp}
\end{center}
\end{figure}
In Fig.2  and 3 the results of the hCQM for the asymptotic behaviour are reported, without and with 
the constituent quark form factors, respectively.
Perturbative QCD (pQCD) predicts \cite{Brodsky} that at high $Q^2$
the helicity conserving Dirac form factor 
$F_{1p} \propto  \frac{1}{Q^4}$ and the helicity-flipping Pauli form factor  
$F_{2p} \propto
\frac{1}{Q^6}$, so that  $Q^2\frac{F_{2p}}{F_{1p}}$ 
should reach a constant value at high enough $Q^2$. 
The asymptotic regime has not been reached yet (see the experimental data).
 Ralston {\it et al.}  showed that if one takes into account the contributions 
from the non zero orbital angular momentum in the proton wave function,
$\frac{F_{2p}}{F_{1p}}$ goes as $\frac{1}{Q}$ \cite{Ralston:2003mt}
 and this kind of scaling behaviour is reported in the right side of Fig.2 and 3.  
We observe that the experimental data for the ratios of Fig. 2 and Fig. 3 are 
quite well reproduced by the theoretical calculations.
As a conclusion we can say that an extension of  the measurement of the form factors 
to higher  $Q^2$ values is important in order to test the pQCD scaling predictions 
for the Dirac and the Pauli form factors  $F_{1p}$ and $F_{2p}$ and in particular to 
understand at which scale this behaviour starts. Moreover, a presence or not of a zero 
in the ratio of the electric and magnetic form factors of the proton, a clearly non 
perturbative effect, will help to discriminate among the different models of the nucleon. 
\begin{figure}[!ht]
\includegraphics[width=4in]{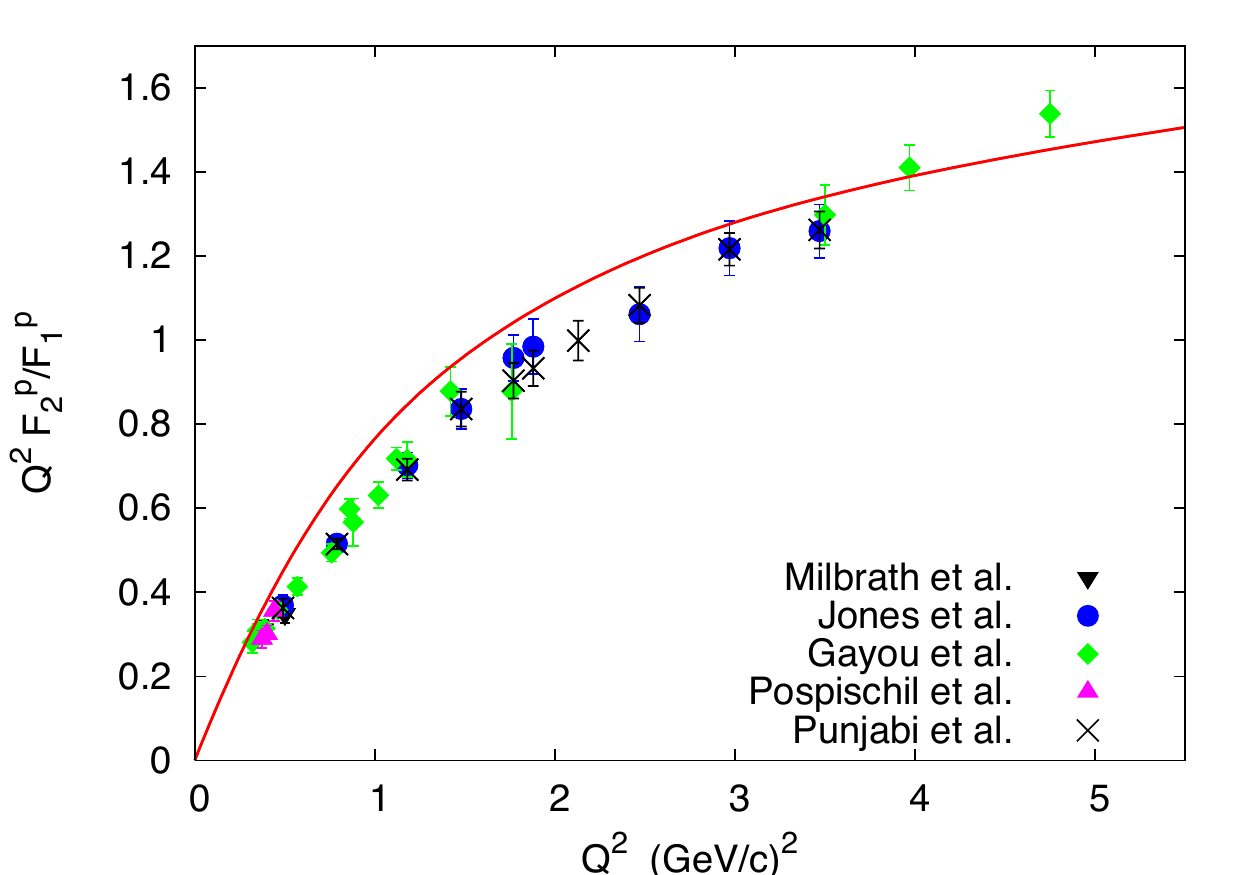}
\includegraphics[width=4in]{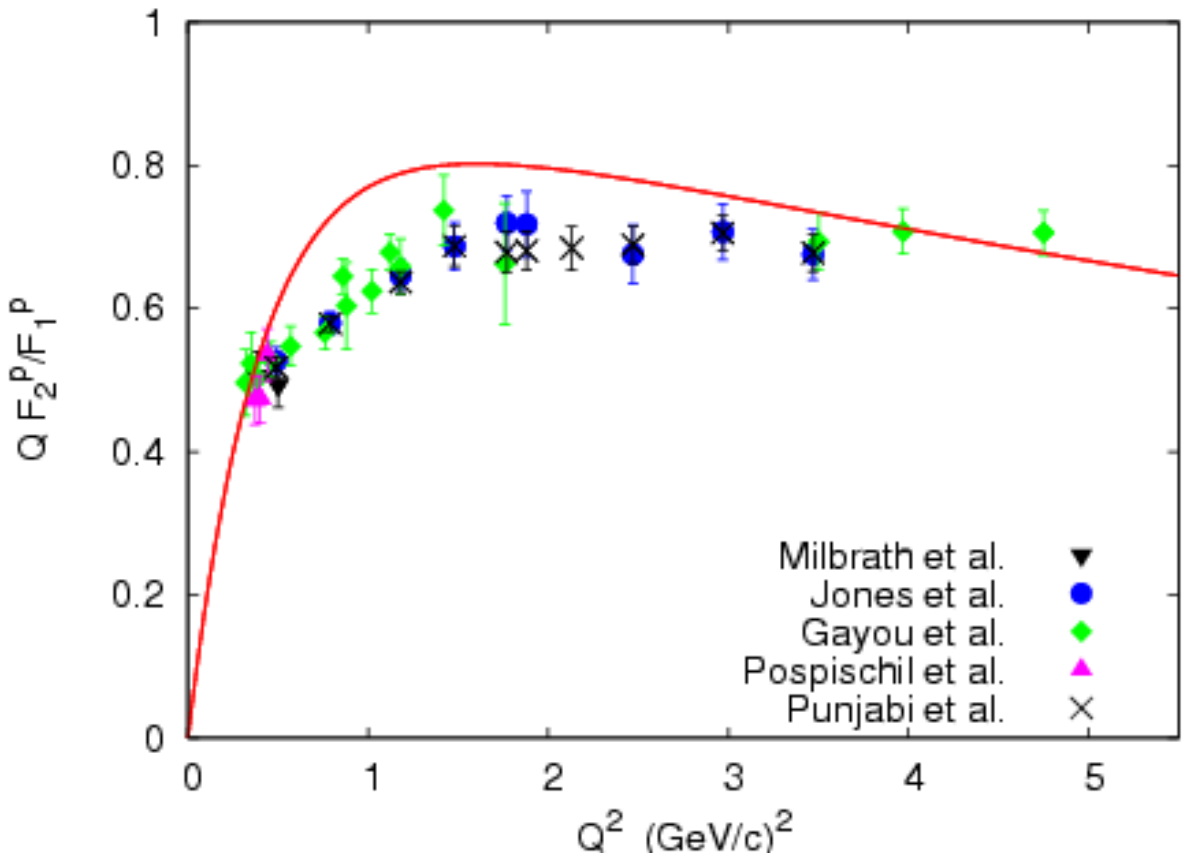}
\caption{(Color on line)
The ratio  $Q^2\frac{F_{2p}}{F_{1p}}$ and $Q\frac{F_{2p}}{F_{1p}}$,  calculated with the relativized hCQM
 (solid line). The experimental data are taken from [19,20,21].} 
\end{figure}
\begin{figure}[!ht]
\includegraphics[width=19pc]{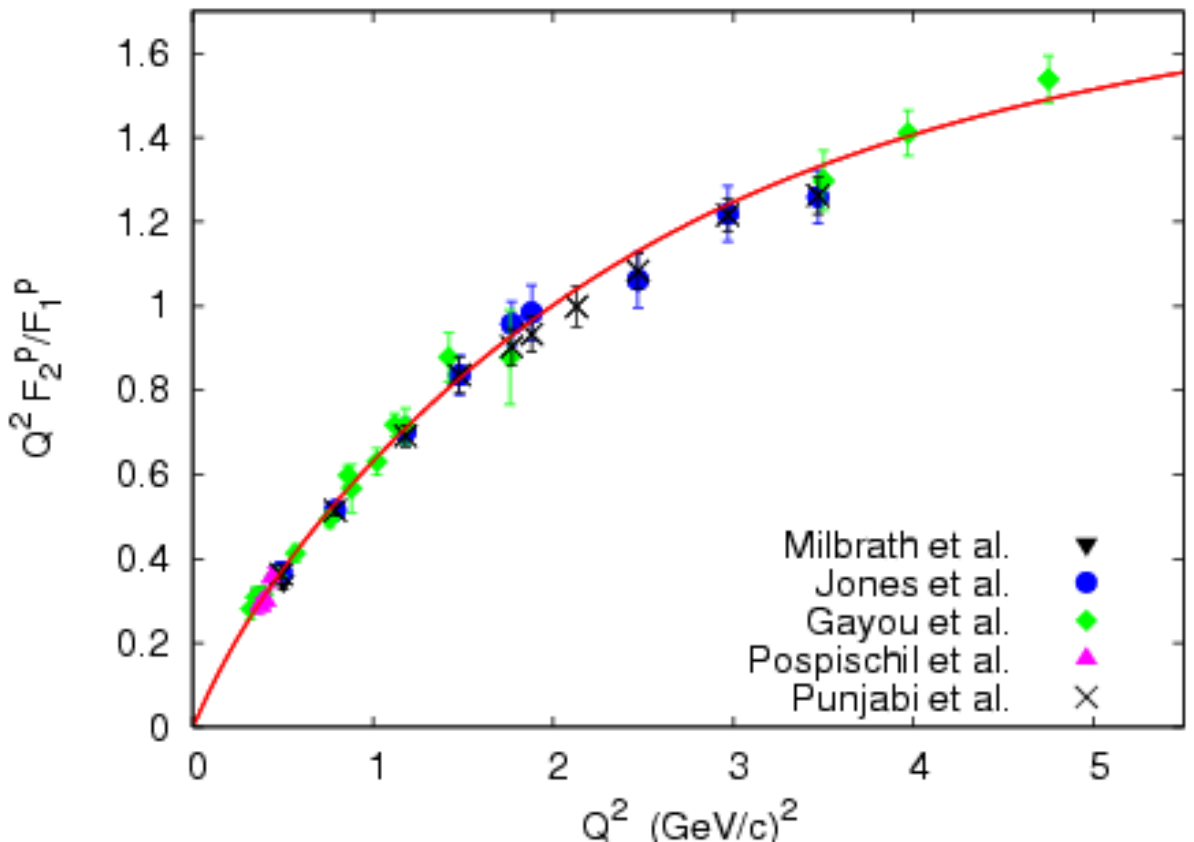}
\includegraphics[width=19pc]{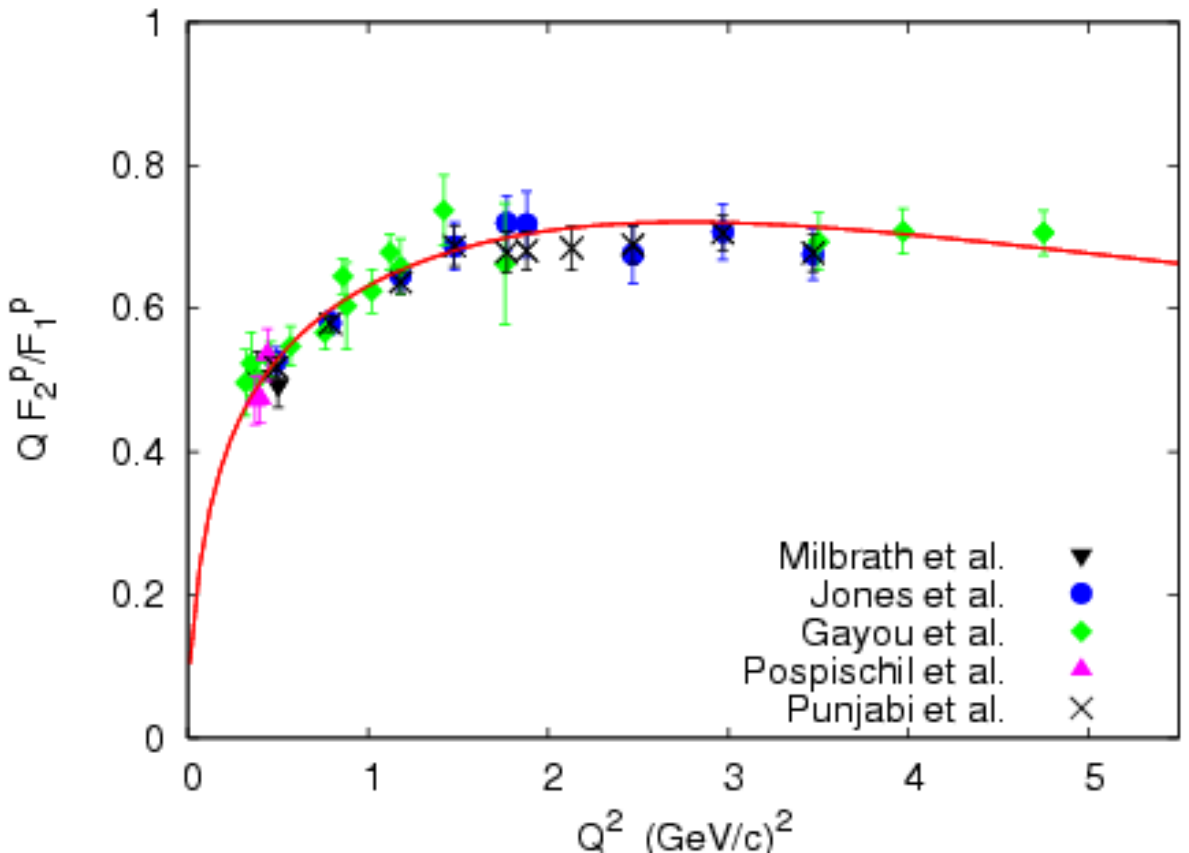}
\caption{(Color on line)The ratio  $Q^2\frac{F_{2p}}{F_{1p}}$ and $Q\frac{F_{2p}}{F_{1p}}$,  calculated with the relativized hCQM with the inclusion of constituent quark form factors
 (solid line). The experimental data are taken from [19,20,21].} 
\end{figure}

\end{document}